\documentclass[12pt]{article}
\usepackage{texdraw}

\begin{document}

\begin{center}
{\Large \bf Integrable potentials on Cayley-Klein spaces \\
 from quantum groups}
\end{center}

\begin{center}
N.A. Gromov, V.V. Kuratov \\
{\it Department of Mathematics, Komi Science Center,\\
 Ural Division, Russian Academy of Sciences, \\
Kommunisticheskaya str., 24, 167982 Syktyvkar, Russia} \\
E-mail: gromov@dm.komisc.ru,  kuratov@dm.komisc.ru
\end{center}


\begin{center}
{ \bf Abstract}
\end{center}

The family of (super)integrable potentials on spaces with curvature 
developed by A. Ballesteros et all is extend    to all two-dimensional Cayley-Klein spaces with the help of contractions. 
It is shown that integrable systems on spaces with degenerate metrics are described by two Hamiltonians: one in the base and another in the fiber.
 
\vspace{5mm}


\section{Introduction}

A family of classical superintegrable systems defined on the two-dimen\-sio\-nal sphere, hyperbolic and (anti) de Sitter  spaces 
was constructed through Hamiltonians defined on the non-standard quantum deformation of a $sl(2)$ Poisson coalgebra \cite{BHR-1,BHR-2,O-92}. All this spaces have a constant curvature that exactly coincides with deformation parameter $z.$ The non-deformed limit
$z\rightarrow 0$ of all these Hamiltonians is then regarded as the zero-curvature limit (contraction) which leads to the corresponding superintegrable systems on the flat Euclidean and Minkowskian spaces. But among  two-dimensional constant curvature Cayley-Klein spaces   there are three spaces with degenerate metric, namely:  flat Galileian ${\bf G}^{1+1}$ and Newtonian ${\bf N}^{1+1}(\pm)$ with non-zero positive and negative curvature. In this paper we modify approach of \cite{BHR-1,BHR-2} in such a way that superintegrable systems are 
defined on all nine two-dimensional  Cayley-Klein spaces. 
We use the method of unified description of Cayley-Klein spaces, groups, algebras etc. \cite{G-2012}.  The main idea of this method is that construction suitable for all  Cayley-Klein cases can be obtained from an analogous construction for spherical space, orthogonal group, orthogonal algebra etc. by an appropriate transformation with the help of contraction parameters.

\section{Quantum group and integrable Hamiltonians}

The non-standard quantum deformation of $sl(2)$ \cite{O-92} written as a Poisson coalgebra $sl_z(2)$ with Poisson bracket and Casimir is given by
\begin{equation}
\{ J_-^*,J^*_+\}=4J_3^*, \quad \{ J_3^*,J^*_+\}=2J_+^*\cosh z^*J_-^*, \quad
\{ J_3^*,J^*_-\}=-2\frac{\sinh z^*J_-^*}{z^*},
\label{1}
\end{equation}
\begin{equation}
 C^*_{z^*}\overline{}=\frac{\sinh z^*J_-^*}{z^*} J_ +^* -J_3^{*2},
\label{2}
\end{equation}
where $z^*$ is a real deformation parameter (we mark initial generators, coordinates, Casimirs etc. by $^*$).
A two-particle symplectic realization of (\ref{1}) in terms of two canonical pairs of coordinates $(q_1,q_2)$ and momenta $(p_1,p_2)$
that depends on two real parameters $b_1,b_2,$ reads \cite{BHR-1,BHR-2}
$$
J_{-}^*=q_2^{*2}+q_1^{*2},\quad
J_3^*=\frac{\sinh z^*q_1^{*2}}{z^*q_1^{*2}}q_1^*p_1^*e^{z^*q_2^{*2}}+
\frac{\sinh z^*q_2^{*2}}{z^*q_2^{*2}}q_2^*p_2^*e^{-z^*q_1^{*2}},
$$
\begin{equation}
J_{+}^*=
\left ( \frac{\sinh z^*q_1^{*2}}{z^*q_1^{*2}} p_1^{*2}+
\frac{z^*b_1^*}{\sinh z^*q_1^{*2}}\right )e^{z^*q_2^{*2}} +
\left ( \frac{\sinh z^*q_2^{*2}}{z^*q_2^{*2}} p_2^{*2}+
\frac{z^*b_2^*}{\sinh z^*q_2^{*2}}\right )e^{-z^*q_1^{*2}}.
\label{3}
\end{equation}
The two-particle Casimir has the form
$$
C^*_{z^*}=\frac{\sinh z^*q_1^{*2}}{z^*q_1^{*2}}\frac{\sinh z^*q_2^{*2}}{z^*q_2^{*2}}(q_1^*p_2^* - q_2^*p_1^*)^2e^{-z^*q_1^{*2}}
 e^{z^*q_2^{*2}} + (b_1^*e^{2z^*q_2^{*2}} + b_2^*e^{-2z^*q_1^{*2}} )
$$
\begin{equation}
+\left (b_1^*\frac{\sinh z^*q_2^{*2}}{\sinh z^*q_1^{*2}} + b_2^*\frac{\sinh z^*q_1^{*2}}{\sinh z^*q_2^{*2}}  \right )
e^{-z^*q_1^{*2}} e^{z^*q_2^{*2}}.
\label{4}
\end{equation}

An arbitrary two-dimensional Cayley-Klein space is obtained from the spherical space by the following transformations of Beltrami coordinates
\begin{equation}
q_1^*=j_1j_2q_1, \quad q_2^*=j_1q_2, 
\label{5}
\end{equation}
where each of parameters $j_k$ takes the values $1,\iota_k,i, \;k=1,2.$ Here $\iota_k$ are nilpotent $\iota_k^2=0$ with commutative law of multiplication $\iota_k \iota_m=\iota_m\iota_k\neq 0, \;k\neq m$ and $\iota_k /\iota_k=1,$ but $\iota_k/\iota_m,\; k\neq m $ or 
$a/\iota_k,\; a\in {\bf R,C}$ are not defined. Nilpotent values of parameters $j_k$ correspond to contractions\footnote{
 In the standard Wigner--In{\"o}n{\"u} contraction procedure \cite{IW} the limit $j_k \rightarrow 0 $ corresponds to the contraction  $j_k=\iota_k$.
}, 
wheares $j_k=i$ correspond to analytical continuations to  pseudoeuclidean spaces. 
The transformations (\ref{5}) induce a transformations of all others constructions,
if additionally to require that the final constructions will be well defined, i.e. do not include nondefined terms like $\iota_k/\iota_m,\; k\neq m $ or $1/\iota_k.$ 
For example, canonical momenta $p_1^*=\frac{\partial}{\partial q_1^*}$ is transformed as follows
$$
p_1=j_1j_2p_1^*(\rightarrow)=j_1j_2\frac{\partial}{\partial j_1j_2q_1}=\frac{\partial}{\partial q_1}.
$$
The arrow $(\rightarrow)$ means that transformed coordinates are substituted instead of initial one according with (\ref{5}).
So both momenta are transformed as 
\begin{equation}
p_1=j_1j_2p_1^*, \quad p_2=j_1p_2^*.
\label{6}
\end{equation}
In what follows the arrow $(\rightarrow)$ will be omitted to simplify notations.

We are now able to transform  generators (\ref{3}) of $ sl_z(2). $
$$
J_{-}=\frac{1}{j_1^2}J^*_{-}(\rightarrow)=q_2^2+j_2^2q_1^2,
$$
$$
J_3=J_3^*(\rightarrow)=\frac{\sinh j_1^2j_2^2zq_1^2}{j_1^2j_2^2zq_1^2}q_1p_1e^{j_1^2zq_2^2}+
\frac{\sinh j_1^2zq_2^2}{j_1^2q_2^2z}q_2p_2e^{-j_1^2j_2^2zq_1^2},
$$
$$
J_{+}=j_1^2j_2^2J_+^*(\rightarrow)=
\left ( \frac{\sinh j_1^2j_2^2zq_1^2}{j_1^2j_2^2zq_1^2}\cdot p_1^2+
\frac{j_1^2j_2^2zb_1}{\sinh j_1^2j_2^2zq_1^2}\right )e^{j_1^2zq_2^2} +
$$
\begin{equation}
+j_2^2\left ( \frac{\sinh j_1^2zq_2^2}{j_1^2zq_2^2}\cdot p_2^2+
\frac{j_1^2zb_2}{\sinh j_1^2zq_2^2}\right )e^{-j_1^2j_2^2zq_1^2},
\label{7}
\end{equation}
We multiply generator $J^*_{-}$ by $j_1^{-2}$ in order to have non-zero generator even for nilpotent values of parameters $j_k.$
Multipliers for generators $J_3^*, J_+^* $ are found from the requirement that the final expressins will be well defined.
It follows that deformation parameter $z^*$ and parameters $b_1^*, b_2^*$ are not transformed
\begin{equation}
z^*=z, \quad b_1^*=b_1, \quad b_2^*=b_2.
\label{8}
\end{equation}
By substituting
$$
J_-^*=j_1^2J_-, \quad J_+^*=\frac{1}{j_1^2j_2^2}J_+, \quad J_3^*=J_3
$$
in (\ref{1}) we obtain Poissson brackets
\begin{equation}
\{J_-,J_+\}=4j_2^2J_3, \quad
\{J_3,J_+\}=2J_+\cosh j_1^2zJ_- , \quad
\{J_3, J_-\}=-2\frac{\sinh j_1^2zJ_-}{j_1^2z}
\label{9}
\end{equation}
of the coalgebra $sl_z(2;j) $ with Casimir
\begin{equation}
 C_z=j_2^2C_z^*(\rightarrow)=\frac{\sinh j_1^2zJ_-}{j_1^2z}\cdot J_+ -j_2^2J_3.
\label{10}
\end{equation}
Two-particle Casimir is obtained from (\ref{4}) by the same transformation and reads
$$
C_{z}=\frac{\sinh j_1^2j_2^2 zq_1^{2}}{j_1^2j_2^2zq_1^{2}}\frac{\sinh j_1^2zq_2^{2}}{j_1^2zq_2^{2}}(j_2^2q_1p_2 - q_2p_1)^2e^{-j_1^2j_2^2zq_1^{2}} e^{j_1^2zq_2^{2}} + j_2^2(b_1e^{2j_1^2zq_2^{2}} + b_2e^{-2j_1^2j_2^2zq_1^{2}} )
$$
\begin{equation}
+ j_2^2\left (b_1\frac{\sinh j_1^2zq_2^{2}}{\sinh j_1^2j_2^2zq_1^{2}} + b_2\frac{\sinh j_1^2j_2^2zq_1^{2}}{\sinh j_1^2zq_2^{2}}  \right )
e^{-j_1^2j_2^2zq_1^{2}} e^{j_1^2zq_2^{2}}.
\label{11}
\end{equation}
This Casimir Poisson-commutes with the generators (\ref{7}) of $sl_z(2;j). $


The simplest integrable and superintegrable Hamiltonians with coalgebra $sl_z(2)$ symmetry introduced in \cite{BHR-1,BHR-2} are
\begin{equation}
 {\cal{H}}_z^{*I}=\frac{1}{2}J_+^*, \quad  {\cal{H}}_z^{*S}=\frac{1}{2}J_+^*e^{z^*J_-^*}
\label{12}
\end{equation}
Both Hamiltonians are proportional to $J_+^* $ therefore are transformed like $J_+^*,$ i.e.
\begin{equation}
 {\cal{H}}_z^I=j_1^2j_2^2{\cal{H}}_z^{*I}(\rightarrow)=\frac{1}{2}J_{+}, \quad \quad
{\cal{H}}_z^S=j_1^2j_2^2{\cal{H}}_z^{*S}(\rightarrow)=\frac{1}{2}J_{+}e^{j_1^2zJ_-}.
\label{13}
\end{equation}

For the flat  spaces $(j_1=\iota_1)$ the  integrable  Hamiltonian and Casimir are given by (\ref{7}),(\ref{11}),(\ref{13})   in the form 
$$ 
 {\cal{H}}_z^I(\iota_1,j_2)=\frac{1}{2}\left( p_1^2+\frac{b_1}{q_1^2} \right) + j_2^2\frac{1}{2}\left(p_2^2+\frac{b_2}{q_2^2} \right),
$$
\begin{equation} 
 C_z(\iota_1,j_2)=q_2^2\left( p_1^2+\frac{b_1}{q_1^2} \right) + j_2^2\left(b_1+b_2-2q_1q_2p_1p_2\right) + j_2^4q_1^2\left(p_2^2+\frac{b_2}{q_2^2}\right).
\label{d1}
\end{equation}
Hamilton equations 
\begin{equation}
 \dot{q}_1=\frac{\partial {\cal{H}}_z^I}{\partial p_1}=p_1, \quad 
 \dot{p}_1=-\frac{\partial {\cal{H}}_z^I}{\partial q_1}=\frac{b_1}{q_1^3}, \quad
 \dot{q}_2=\frac{\partial {\cal{H}}_z^I}{\partial p_2}=j_2^2 p_2, \quad
 \dot{p}_2=-\frac{\partial {\cal{H}}_z^I}{\partial q_2}=j_2^2 \frac{b_2}{q_2^3}
\label{d2}
\end{equation}
are easily solved
\begin{equation}
 q_1^2(t)=\frac{b_1}{E_1}+E_1\left(t-t_0\right)^2, \quad
  q_2^2(t)=\frac{b_2}{E_2}+j_2^4E_2\left(t-t_0\right)^2,
\label{d3}
\end{equation}
where
\begin{equation}
E_1 = \dot{q}_1^2 + \frac{b_1}{q_1^2}, \quad
j_2^4E_2 = \dot{q}_2^2 +j_2^4 \frac{b_2}{q_2^2}
\label{d4}
\end{equation}
are constants of the motion and $t_0$ is an integration constant.
Parametric solution (\ref{d3}) represets the following trajectory 
\begin{equation}
  E_1q_2^2- j_2^4E_2q_1^2=b_2\frac{E_1}{E_2}-j_2^4b_1\frac{E_2}{E_1},
\label{d5}
\end{equation}
which is  hyperbola for Euclidean  ${\bf E}^2, (j_2=1)$ and Minkowskian ${\bf M}^{1+1}, (j_2=i)$ planes and is contracted to the one-dimensional fiber
\begin{equation} 
F_1(q_2^0)=\left\{ q_1^2(t)=\frac{b_1}{E_1}+E_1\left(t-t_0\right)^2, \quad 
 q_2=\pm\sqrt{b_2/E_2}=q_2^0=const \right\}
\label{d6}
\end{equation}
for Galilean plane ${\bf G}^{1+1}, (j_2=\iota_2)$ with the 
base along the Cartesian coordinate $q_2$. This motion is defined by the non-zero part of Hamiltonian (\ref{d1})
\begin{equation}
{\cal{H}}_{z,f}^I(\iota_1,\iota_2)=\frac{1}{2}\left( p_1^2+\frac{b_1}{q_1^2} \right).
\label{d6-1}
\end{equation}
The motion in the base is independent on the motion in the fiber and is defined by the second part 
($\approx j_2^2$) of  ${\cal{H}}_z^I(\iota_1,j_2)$ (\ref{d1})
\begin{equation}
{\cal{H}}_{z,b}^I(\iota_1,\iota_2)=\frac{1}{2}\left(p_2^2+\frac{b_2}{q_2^2} \right),
\label{d6-2}
\end{equation}
which gives the following trajectory 
\begin{equation}
B_2=\left\{  q_2^2(\tau)=\frac{b_2}{\hat{E}_2}+\hat{E}_2\left(\tau-\tau_0\right)^2, \quad
\hat{E}_2 = \left(\frac{dq_2}{d\tau}\right)^2 + \frac{b_2}{q_2^2}=const \right\}.
\label{d6-3}
\end{equation}
We introduce new variable $\tau$ instead of $t$ in order to stress independence of base and fiber motions.

Equations (\ref{d6}) and (\ref{d6-3}) illustrate the general properties of physical system with non-semisimple symmetry  group \cite{G-2012}. Let physical system has a simple or semisimple  symmetry  group $G$. The operation of group contraction transforms   $G$ to a non-semisimple group with the structure of a semidirect product $G=A{\times \!\!\!\!\!\! \supset} G_1$, where $A$ is Abel and $G_1\subset G$ is an untouched subgroup. At the same time the  representation space of the group $G$ is fibered under   the contraction in such a way that   the subgroup $G_1$ acts in the fiber.
The simple and  the best known example is Galilei group 
$G(1,3)=T_4{\times \!\!\!\!\!\! \supset}SO(1,3)$ and  the  non-relativistic space--time ${\bf G}^{1+3}$ with  has one-dimensional base which is interpreted as  time, and  three-dimensional fiber, which is interpreted as proper space.
 Contraction of the symmetry  group correspond to some limit case of the physical system, which is divided on two subsystems: one in the base $S_b$ and  the other subsystem $S_f$ in the  fiber.
$S_b$ forms a closed system since according to semi-Riemannian geometry \cite{P-65-1,G-09}  the properties of the base do not depend on  the points of the fiber, which physically means    that  the subsystem $S_f$ 
have not effect   on the $S_b$.
On the contrary  the properties of the fiber depend on  the points of the base, therefore the subsystem $S_b$
exerts influence upon $S_f$. More precisely,  $S_b$ specify  outer (or background) conditions for $S_f$
 in every fiber.


The second particular case is given by the constant curvature Newtonian spaces ${\bf N}^{1+1}(\pm), (j_1=1,i)$ with degenerate metric $(j_2=\iota_2)$. The  integrable  Hamiltonian and Casimir for the motion in the fiber are obtained from (\ref{7}),(\ref{11}),(\ref{13}) in the form
\begin{equation}  
{\cal{H}}_{z,f}^I(j_1,\iota_2)=\frac{1}{2}\left( p_1^2+\frac{b_1}{q_1^2} \right)e^{j_1^2zq_2^2},\quad
C_{z,f}(j_1,\iota_2)=\frac{\sinh j_1^2zq_2^2}{j_1^2zq_2^2}\left( p_1^2+\frac{b_1}{q_1^2} \right)e^{j_1^2zq_2^2}.
\label{d7}
\end{equation}
Hamiltonian ${\cal{H}}_{z,f}^I(j_1,\iota_2)$ does not depend on the  momenta $p_2$, therefore
 equation of motion for the second coordinate have the form $\dot{q}_2=0$ with the solution $q_2=q_2^0=const.$
Then Hamilton equations for the first canonical pair take the form
\begin{equation}
\dot{q}_1=\frac{\partial {\cal{H}}_{z,f}^I}{\partial p_1}=p_1e^{j_1^2z(q_2^0)^2}, \quad 
 \dot{p}_1=-\frac{\partial {\cal{H}}_{z,f}^I}{\partial q_1}=\frac{b_1}{q_1^3}e^{j_1^2z(q_2^0)^2}
\label{d8}
\end{equation}
with the solution
\begin{equation}
 q_1^2(t)=\frac{b}{E_1}+E_1\left(t-t_0\right)^2,
\label{d9}
\end{equation}
where
\begin{equation}
E_1 = \dot{q}_1^2 + \frac{b}{q_1^2}, \quad b=b_1e^{2j_1^2z(q_2^0)^2}
\label{d10}
\end{equation}
is constant of the motion and $t_0$ is an integration constant.
So for nonzero curvature the trajectory  (\ref{d9}) belong to the fiber $q_2=q_2^0$ as for Galilei space
 ${\bf G}^{1+1},$ but  depend on the fiber through the effective barrier parameter $b$ (\ref{d10}).
The motion in the base is defined by the Hamiltonian ${\cal{H}}_{z,b}^I(j_1,\iota_2)$ which consists of proportional to $j_2^2$ terms in  (\ref{7}) and has the form
\begin{equation}  
{\cal{H}}_{z,b}^I(j_1,\iota_2)=\frac{1}{2}\left ( \frac{\sinh j_1^2zq_2^2}{j_1^2zq_2^2}\cdot p_2^2+
\frac{j_1^2zb_2}{\sinh j_1^2zq_2^2}\right ).
\label{d10-1}
\end{equation}

In the broad sense of the word  deformation is inverse operation to contraction. The non-trivial deformation
of some algebraic structure generally means 
its non-evident generalization. Quantum groups \cite{FRT}, which are simultaneously non-commutative and non-cocommutative Hopf algebras, present a good example of similar generalization since previously only commutative and non-cocommutative or non-commutative and cocommutative Hopf algebras was known.
But when   contraction of  some mathematical structure is performed one can reconstruct  the initial structure by the deformation in the narrow sense moving back along the contraction way.
Similar approach was used to describe the early history of the Universy starting from the electroweak model \cite{G-15}.  Hamiltonians in the base ${\cal{H}}_{z,b}^I(j_1,\iota_2)$ (\ref{d10-1}) and ${\cal{H}}_{z,b}^I(\iota_1,\iota_2)$ (\ref{d6-2}) are obtained namely in this way.

The same law of transformation is hold for the integrable Smorodinsky-Winternitz (SW) and Kepler-Coulomb (KC) potentials
\begin{equation}
 {\cal{H}}_z^{*ISW}=\frac{1}{2}J_+^*  + \beta_0^*\frac{\sinh z^*J_-^*}{z^*}, \quad
 {\cal{H}}_z^{*IKC}=\frac{1}{2}J_+^* - \gamma^*\sqrt{\frac{2z^*}{e^{2z^*J_-^*}-1}}e^{2z^*J_-^*}.
\label{14}
\end{equation}
Taking into consideration  that  for Cayley-Klein spaces with degenerate metric both potential depend only on base variable and  therefore must appear among base terms we obtain  integrable  Hamiltonians 
in the form
$$
 {\cal{H}}_z^{ISW}=j_1^2j_2^2{\cal{H}}_z^{*ISW}(\rightarrow)=\frac{1}{2}J_{+} +j_2^2\beta_0\frac{\sinh j_1^2zJ_-}{j_1^2z},
$$
\begin{equation}
 {\cal{H}}_z^{IKC}=j_1^2j_2^2{\cal{H}}_z^{*IKC}(\rightarrow)=\frac{1}{2}J_{+} -j_2^2\gamma
\sqrt{\frac{j_1^22z}{e^{j_1^22zJ_-}-1}}e^{j_1^22zJ_-},
\label{15}
\end{equation}
where transformation laws for the constants $\beta_0^*, \gamma^*$  are given by 
\begin{equation}
\beta_0=j_1^4\beta_0^*, \quad \gamma =j_1\gamma^*.
\label{16}
\end{equation}
The  superintegrable Hamiltonian with the SW potential look as follows
\begin{equation}
 {\cal{H}}_z^{SSW}=j_1^2j_2^2{\cal{H}}_z^{*SSW}(\rightarrow)=\frac{1}{2}J_{+}e^{j_1^2zJ_-} +j_2^2\beta_0\frac{\sinh j_1^2zJ_-}{j_1^2z}e^{j_1^2zJ_-}
\label{15-1}
\end{equation}
and analogous expression for KC potential.


\section{Polar coordinates on Cayley-Klein spaces }  

In \cite{BHR-1,BHR-2} new coordinates $\rho^*, \theta^* $ are introduced by relations
\begin{equation}
 \cosh\rho^*=\exp \left \{z^*(q_1^{*2}+q_2^{*2})\right \}\equiv e^{z^*J_-^*}, \quad
 \sin^2\theta^*=\frac{1 - \exp \left \{2z^*q_1^{*2}\right \}}{1 - \exp\{2z^*(q_1^{*2}+q_2^{*2})\}}
\label{17}
\end{equation}
and the metric as well as Gaussian curvature are written as
\begin{equation}
ds^{*2}=\frac{1}{\cosh \rho^* }(d\rho^{*2} + \sinh^2\rho^*d\theta^{*2}), \quad
K^*(\rho^*)=-\frac{z^*\sinh^2\rho^*}{2\cosh\rho^*}.  
\label{18}
\end{equation}
The product $\cosh (\rho^*)ds^{*2}$ coincides with the metric of the two-dimensional space with constant curvature $ \kappa =-z^*$
provided that $ (\rho^*, \theta^*)$ are proportional to geodesic polar coordinates. To obtain the spherical space we take $z^*=-1,$
so that $ \kappa =1.$ In the limit $z\rightarrow 0$ we have from the first equation of (\ref{17}) 
$\rho^{*2}=2z^*(q_2^{*2} + q_1^{*2})=-2(q_2^{*2} + q_1^{*2}),$ i.e.
$\rho^{*}=i\sqrt{2(q_2^{*2} + q_1^{*2})}.$
If one introduce coordinates $x^*=\sqrt{2}q_2^*, \, y^*=\sqrt{2}q_1^* $ and 
  $r^*=\sqrt{x^{*2}+y^{*2}}, $ then $\rho^*=ir^*. $
We assume this relation for arbitrary $z,$ i.e. radial coordinate $r^*$   is defined as
\begin{equation}
 \cosh r^*=\exp \left \{-(q_1^{*2}+q_2^{*2})\right \}=\exp \left \{-\frac{1}{2}(x^{*2}+y^{*2})\right \}.
 \label{19}
\end{equation}
The metric and  Gaussian curvature (\ref{18})  are rewritten in  polar coordinates $(\rho^*, \theta^{*}) $ as
\begin{equation}
ds^{*2}=\frac{1}{\cos r^*}\left (dr^{*2} + \sin^2r^* d\theta^{*2} \right), \quad
K^*(r^*)=-\frac{\sin^2r^*}{2\cos r^*}.
\label{20}
\end{equation}
The metric (\ref{20}) correspond to the isotropic space with non-constant curvature that depend only on radial coordinate $r^*.$
The metric $d\tilde{s}^{*2}=\cos r^*ds^{*2}$ describe spherical space with constant curvature. Therefore we can introduce contraction parameters $j_1, j_2$ as usual.

It is easily to obtain the transformation laws for Beltrami $ (x^*, y^*)$ and polar  $(r^*, \theta^{*}) $ coordinates, namely
\begin{equation}
x^*=j_1x, \;\; y^*=j_1j_2y, \quad r^*=j_1r, \;\; \theta^*=j_2\theta.  
\label{21}
\end{equation}
Relations of Beltrami $ (x, y)$ and polar  $(r, \theta) $ coordinates for all Cayley-Klein cases are
\begin{equation}
\cos j_1r=\exp \left \{-\frac{1}{2}j_1^2(x^2+j_2^2y^2)\right \}, \quad
\frac{1}{j_2^2}\sin^2 j_2\theta
=\frac{1}{j_2^2}\frac{1 - \exp \left \{-j_1^2j_2^2y^2\right \}}{1 - \exp \left \{-j_1^2(x^2+j_2^2y^2)\right \}}.
\label{22}
\end{equation}
The metric  is given by
\begin{equation}
ds^2=\frac{1}{j_1^2}ds^{*2}(\rightarrow)=\frac{1}{\cos j_1r}\left (dr^2+j_2^2\frac{\sin^2j_1r}{j_1^2} d\theta^2 \right). 
\label{23}
\end{equation}
For nilpotent value of parameter $j_2=\iota_2,$ i.e. for fiber Newtonian  $N^{1+1}(\pm), (j_1=1,i)$ and  Galileian 
$G^{1+1},(j_1=\iota_1)$ spaces  this metric is degenerate. In fact for fiber spaces there are two metrics: one for the base and another for the fiber. In the polar coordinates metrics is represented by the  radial ($\sim$ base) and the angle ($\sim$ fiber) parts
\begin{equation}
ds_r^2=\frac{1}{\cos j_1r}dr^2, \quad  ds_{\theta}^2=\frac{\sin^2j_1r}{j_1^2\cos j_1r} d\theta^2, 
\label{24}
\end{equation}
which looks for the flat Galilei space as $ds_r^2=dr^2, \; ds_{\theta}^2=r^2d\theta^2.$

\section{The superintegrable  potentials on Cayley-Klein spaces }  

Let $(p^*_{r}, p^*_{\theta})$ be the canonical momenta corresponding to the new polar coordinates $(r^*, \theta^*). $
Transformation laws of these momenta follow from (\ref{21}) in the form
\begin{equation}
p_{r}=j_1p^*_{r}, \quad p_{\theta}=j_2p^*_{\theta}.
\label{25}
\end{equation}
The generic integrable Hamiltonians (3.3) \cite{BHR-2} after substitution $\rho=ir^*,  p_\rho=-ip^*_{r}, \lambda_1=\lambda_2=1  $
takes the form
\begin{equation}
H_z^{*I}=\frac{1}{2}\cos r^* \left( p^{*2}_{r} +\frac{1}{\sin^2r^*} p^{*2}_{\theta} \right)  +
\frac{2\cos r^*}{\sin^2 r^*}\left(\frac{b_1^*}{\sin^2\theta^*} +
\frac{b_2^*}{\cos^2\theta^*}\right)+ g^*(r^*)
\label{26}
\end{equation}
and transforms with the help of (\ref{8}), (\ref{13}), (\ref{21}), (\ref{25}) to all Cayley-Klein spaces
$$
H_z^I=j_1^2j_2^2H_z^{*I}(\rightarrow)=
$$
$$
=\frac{1}{2}\cos j_1r
\left( j_2^2p^2_r+\frac{j_1^2}{\sin^2j_1r}p^2_\theta \right)+
\frac{2j_1^2\cos j_1r}{\sin^2j_1r}
\left( \frac{j_2^2b_1}{\sin^2j_2\theta}+j_2^2\frac{b_2}{\cos^2j_2\theta}\right )+ j_2^2g(j_1r)=
$$
\begin{equation}
=j_2^2\frac{1}{2}\cos j_1r p^2_r + \frac{j_1^2\cos j_1r}{2\sin^2j_1r}C_z + j_2^2g(j_1r),
\label{27}
\end{equation}
where $g(j_1r)=j_1^2g^*(\rightarrow)$.
The corresponding constant of the motion is given by Casimir
\begin{equation}
 C_z(j_1,j_2)=4j_2^2C_z^*(\rightarrow)=p^2_\theta +\frac{4j_2^2b_1}{\sin^2j_2\theta}+
j_2^2\frac{4b_2}{\cos^2j_2\theta}
\label{28}
\end{equation}
which coincides with (3.4) in \cite{BHR-2} and does not depend on $(r, p_r).$
These expressions for Hamiltonians (\ref{27}) and Casimirs (\ref{28}) coincide with the corresponding expressions of Table 3 in \cite{BHR-2}
for deformed sphere ${\bf S}_z^2, (j_1=j_2=1),$ deformed Lobachevsky (or hyperbolic) space ${\bf H}_z^2, (j_1=i, j_2=1),$ deformed anti-de Sitter space-time 
${\bf AdS}_z^{1+1}, (j_1=1, j_2=i),$ deformed de Sitter space-time ${\bf dS}_z^{1+1}, (j_1=i, j_2=i),$ Euclidean space ${\bf E}^2, (j_1=\iota_1, j_2=1),$
Minkowskian space-time ${\bf M}^{1+1}, (j_1=\iota_1, j_2=i),$ but moreover provide the expressions for spaces with degenerate metric:
deformed Newtonian ${\bf N}^{1+1}(\pm), (j_2=\iota_2),$ $j_1=1$ -- positive curvature,  $j_1=i$ -- negative curvature and flat Galileian ${\bf G}^{1+1},(j_1=\iota_1, j_2=\iota_2).$ 
For anti-de Sitter, de Sitter and Minkowskian space-time Hamiltonian $H_z^I $ need be modify for $-H_z^I $.

Just as the metrics (\ref{23}) for fiber spaces is represented by the radial and the angle parts (\ref{24}),
Hamiltonian (\ref{27}) for Newtonian  ${\bf N}^{1+1}(\pm)$ and Galilean ${\bf G}^{1+1}$ spaces
 is divided on two Hamiltonians.
In particular, for $j_2=\iota_2$ equation (\ref{27}) gives  radial (base) and angle (fiber) Hamiltonians
$$
H_r^I(j_1,\iota_2)= \left (\frac{1}{2}p_r^2 + \frac{2j_1^2b_2}{\sin^2j_1r} +\frac{j_1^2g(j_1r)}{\cos j_1r} \right )\cos j_1r = H_r(j_1,\iota_2)\cos j_1r,
$$ 
\begin{equation}
H_{\theta}^I(j_1,\iota_2)=j_1^2\frac{\cos j_1r}{2\sin^2j_1r}p^2_\theta  +
\frac{2j_1^2b_1\cos j_1r}{\theta^2\sin^2j_1r}= H_{\theta}(j_1,\iota_2)\cos j_1r =
j_1^2\frac{\cos j_1r}{2\sin^2j_1r}C_z(\iota_2)
\label{29}
\end{equation}
and  for Galileian space ${\bf G}^{1+1}$  looks as follows 
\begin{equation}
H_r^I(\iota_1,\iota_2)= \frac{1}{2}p_r^2 + \frac{2b_2}{r^2}+ g(r), \quad
H_{\theta}^I(\iota_1,\iota_2)=\frac{1}{2r^2}p^2_\theta  +\frac{2b_1}{\theta^2r^2}=\frac{1}{2r^2}C_z(\iota_2),
\label{30}
\end{equation}
where in both cases the constant of motion (\ref{28}) is equal to
\begin{equation}
C_z(\iota_2)=p^2_\theta +\frac{4b_1}{\theta^2}.
\label{31}
\end{equation}

Superintegrable Hamiltonian $H_z^*$  on constant curvature spherical space is related with integrable Hamiltonian $H_z^{*I}$ by \cite{BHR-2}
\begin{equation}
H_z^{*I}= H_z^{*}\cos r^*.
\label{32}
\end{equation}
The potential functions $g^*(r^*)$ appearing in (\ref{26}) that correspond to Smorodinsky-Winternitz $H_z^{*SW}$ and Kepler-Coulomb  $H_z^{*KC}$ Hamiltonians
(\ref{15}) reads 
\begin{equation}
g^*(r^*)=\beta^*_0\cos r^*\tan^2r^*, \quad g^*(r^*)=-\frac{k^*\cos r^*}{\tan r^*},
\label{33}
\end{equation}
where $k^*=i2\sqrt{2}\gamma^*.$ Taking into account the transformation laws (\ref{15}), (\ref{16}), we obtain Smorodinsky-Winternitz $H_z^{SW}$ and Kepler-Coulomb  $H_z^{KC}$ Hamiltonians for all Cayley-Klein spaces of constant curvature in the form
\begin{equation}
H_z^{SW}=\frac{1}{2}\left (j_2^2p_r^2+\frac{j_1^2}{\sin^2j_1r}p^2_\theta \right)
 + \frac{2j_1^2}{\sin^2j_1r}
\left( \frac{j_2^2b_1}{\sin^2j_2\theta}+j_2^2\frac{b_2}{\cos^2j_2\theta} \right)+ j_2^2\beta_0\frac{\tan^2j_1r}{j_1^2},
\label{34}
\end{equation}
\begin{equation}
H_z^{KC}=\frac{1}{2}\left (j_2^2p^2_r+
\frac{j_1^2}{\sin^2j_1r}p^2_\theta \right)  +
\frac{2j_1^2}{\sin^2j_1r}\left (
\frac{j_2^2b_1}{\sin^2j_2\theta}+j_2^2\frac{b_2}{\cos^2j_2\theta}\right)- \frac{j_2^2j_1k}{\tan j_1r}.
\label{35}
\end{equation}
Again these Hamiltonians are identical with those in Table 4 \cite{BHR-2} for spaces with non-degenerate metric.
For  Newton spaces ${\bf N}^{1+1}(\pm)$ with non-zero curvature expressions (\ref{34}),(\ref{35}) give the base (radial) Hamiltonians
\begin{equation}
H_{r}^{SW}(j_1,\iota_2)=\frac{1}{2}\left (p^2_r +  
\frac{j_1^24b_2}{\sin^2j_1r}\right ) + \beta_0\frac{\tan^2j_1r}{j_1^2},
\label{36}
\end{equation}
\begin{equation}
H_{r}^{KC}(j_1,\iota_2)=\frac{1}{2}\left (p^2_r +  
\frac{j_1^24b_2}{\sin^2j_1r}\right ) - \frac{j_1k}{\tan j_1r}.
\label{36-1}
\end{equation}
The fiber (angle) Hamiltonians are identical in both cases
\begin{equation}
H_{\theta}^{SW}(j_1,\iota_2)=H_{\theta}^{KC}(j_1,\iota_2)=\frac{j_1^2}{2\sin^2j_1r}\left (p^2_\theta +  
\frac{4b_1}{\theta^2}\right ).
\label{37}
\end{equation}
For Galilei space ${\bf G}^{1+1}$ these Hamiltonians looks as follows
$$
H_{r}^{SW}(\iota_1,\iota_2)=\frac{1}{2}\left ( p^2_r  + \frac{4b_2}{r^2}\right)+ \beta_0r^2, \quad
H_{r}^{KC}(\iota_1,\iota_2)=\frac{1}{2}\left ( p^2_r  + \frac{4b_2}{r^2}\right)- \frac{k}{r},
$$
\begin{equation}
H_{\theta}^{SW}(\iota_1,\iota_2)=H_{\theta}^{KC}(\iota_1,\iota_2)=\frac{1}{2r^2}\left ( p^2_\theta  + \frac{4b_1}{\theta^2}\right)
\label{38}
\end{equation}
and represent the different motions in the base with the same motion in the fiber. 

\section{Conclusion}

Starting from the explicit expressions  obtained in \cite{BHR-1,BHR-2} for
deformed sphere ${\bf S}_z^2$ we derive a general expressions of integrable and superintegrable Hamiltonians and corresponding Casimirs suitable for all Cayley-Klein spaces including those with degenerate metric, which   geometry is semi-Riemanian one with one-dimensional base and one-dimensional fiber.
In the last case the  whole motion of the system is divided on two independent motions: one in the base and other in the fiber, which are described by two Hamiltonians ${\cal H}_b$ and ${\cal H}_f$. 

We have demonstrate this limit process on the simple contraction of flat Euclidean space  to Galilean space-time when hyperbolic trajectory is transformed 
to the straight line fiber.
There is little point in speculating about the base and fiber motions
if only Hamiltonian (\ref{d6-1}) is known. But situation is quite different in the case of contraction, 
where for integrable Hamiltonian (\ref{d1})  
 its base part (\ref{d6-2}) 
 tends to zero and can be reconstracted with the help of  deformation which is performed   back along the contraction way. The same is held for Hamiltonians (\ref{d7}) and (\ref{d10-1}) of Newtonian spaces ${\bf N}^{1+1}(\pm)$.
The notion of semi-Riemanian geometry with base and fiber motions is the way to keep all information on the contracted system.

\end{document}